\documentclass[aps,prl,twocolumn,preprintnumbers,10pt,showpacs]{revtex4-1}
\usepackage{epsfig,psfrag,amsmath,amssymb,lmodern}
\usepackage[normalem]{ulem}
\usepackage{color}
\usepackage[applemac]{inputenc}

\usepackage[T2A,T1]{fontenc}

\def \e  {\mathop{\rm e}\nolimits}
\newcommand\lr[1]{{\left({#1}\right)}}

\newcommand{\cN}{{\cal N}}

\newcommand{\cG}{{\cal G}}

\newcommand{\nt}{\notag\\} 
\newcommand{\z}{\zeta}

\renewcommand{\a}{\alpha}
\newcommand{\ep}{\epsilon}
\newcommand{\bz}{\bar z}
\newcommand{\lc}{\circlearrowleft}
\newcommand{\rc}{\circlearrowright}
 
\newcommand{\p}[1]{(\ref{#1})}

\newcommand \vev [1] {\langle{#1}\rangle}

\newcommand{\EEC}{\rm EEC}

\def \be  {\begin{equation}}
\def \ee  {\end{equation}}
\def \ba  {\begin{eqnarray}}
\def \ea  {\end{eqnarray}}
\def \baa {\begin{eqnarray*}}
\def \eaa {\end{eqnarray*}}
\def \bb  {\begin {thebibliography} }
\def \eb  {\end{thebibliography}}

\begin{document}
\title{
\begin{flushright} ${}$\\[-40pt] $\scriptstyle \rm CERN-TH-2019-026,\ LAPTH-014/19,\ MPP-2019-54 $ \\[0pt]
\end{flushright}
Energy-energy correlations at next-to-next-to-leading order
}

\author{J.\ M.\ Henn$^{a}$, E.\ Sokatchev$^{b}$, K.\ Yan$^{a}$, A.\ Zhiboedov$^{c}$}

\affiliation{
$^a$ Max-Planck-Institut f{\"u}r Physik, Werner-Heisenberg-Institut, 80805 M{\"u}nchen, Germany\\
$^b$ LAPTh, Universit\'e Savoie Mont Blanc, CNRS, B.P. 110, F-74941 Annecy-le-Vieux, France\\
$^c$ Theoretical Physics Department, CERN, 1211 Geneva 23, Switzerland}

\begin{abstract}
We develop further an approach to computing energy-energy correlations (EEC) directly from finite correlation functions. In this way, one completely avoids infrared divergences. In maximally supersymmetric Yang-Mills theory ($\mathcal{N}=4$ sYM), we derive a new, extremely simple formula relating the EEC to a triple discontinuity of a four-point correlation function. We use this formula to compute the EEC in $\mathcal{N}=4$ sYM at next-to-next-to-leading order in perturbation theory. Our result is given by a two-fold integral representation that is straightforwardly evaluated numerically. We find that some of the integration kernels are equivalent to those appearing in sunrise Feynman integrals, which evaluate to elliptic functions. 
Finally, we use the new formula to provide the expansion of the EEC in the back-to-back and collinear limits. 
\end{abstract}

\maketitle

\begin{figure}[t]
\includegraphics[width = 0.37\textwidth]{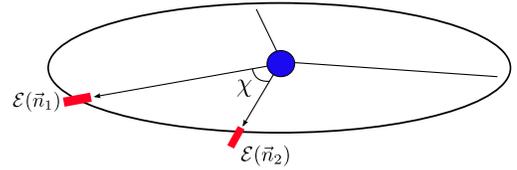}
\caption{\small Graphical representation of the energy-energy correlation: particles produced out of the vacuum by
the source are captured by the two detectors located at spatial infinity in the directions of the unit vectors $\vec n_1$ and $\vec n_2$.
}
\label{fig-EE}
\end{figure}

\noindent \textbf{1. Introduction. }  
The energy-energy correlation (EEC) \cite{Basham:1978bw} measures the energy flow through a pair of detectors separated by an angle $\chi$, see Fig.~\ref{fig-EE}. The EEC has several nice properties, and may help to understand better the nature of jets in quantum field theory. 
It is an infrared-safe observable \cite{Kinoshita:1962ur,Lee:1964is} that can be computed perturbatively.
Moreover, it has simple factorization properties in the back-to-back ($\chi \to \pi$) \cite{Collins:1981uk,Kodaira:1981nh,deFlorian:2004mp,Tulipant:2017ybb,Moult:2018jzp} and collinear ($\chi \to 0$) \cite{Konishi:1978yx,Konishi:1978ax} limit. This knowledge can be used to match fixed-order predictions to resummation calculations \cite{deFlorian:2004mp,Tulipant:2017ybb,Moult:2018jzp}.
On the other hand, the EEC is experimentally measurable, and in particular has been used for precision tests of QCD and measurement of the strong coupling constant  $\alpha_s$ \cite{Aaboud:2017fml,Kardos:2018kqj,Verbytskyi:2019zhh}.

The EEC at leading order (LO) is known since \cite{Basham:1978bw},
while the next-to-leading order (NLO) and next-to-next-to-leading order (NNLO) results were obtained numerically in refs. \cite{Richards:1982te,Richards:1983sr,Ali:1982ub,Falck:1988gb,Kunszt:1989km,Glover:1994vz,Clay:1995sd,Kramer:1996qr} and \cite{DelDuca:2016csb,DelDuca:2016ily}, respectively.
Only very recently, the NLO result was computed analytically \cite{Dixon:2018qgp}.

Although the EEC is infrared finite, the standard approach to computing
it involves infrared divergent scattering amplitudes \cite{Gituliar:2017umx,Dixon:2018qgp}.
On the other hand, it can be defined starting from correlation functions, which are 
infrared finite \cite{Sveshnikov:1995vi,Korchemsky:1997sy,Korchemsky:1999kt,Belitsky:2001ij,Hofman:2008ar}. 
For example, for $e^+ + e^- \rightarrow  \gamma^* \rightarrow X$, the main ingredient is a four-point correlation function of two energy-momentum tensors (representing the two detectors), and two electromagnetic currents, which create the electron-positron pair from the vacuum. 

To the best of our knowledge, this approach has not yet been implemented in QCD. 
On the other hand, these ideas were applied in $\mathcal{N}=4$ super Yang-Mills (sYM)  \cite{Hofman:2008ar,Belitsky:2013xxa,Belitsky:2013bja}, culminating in the first analytic calculation of and EEC at NLO \cite{Belitsky:2013ofa}.
The structure of this result, and in particular the types of polylogarithmic functions appearing in it, foreshadowed the structures later found in QCD \cite{Dixon:2018qgp}.

We show that for an analog of the electromagnetic current in $\mathcal{N}=4$ sYM,
the EEC is computed by a new, extremely simple formula, 
given by a two-fold integral of a particular \textit{triple discontinuity} of the four-point correlation function, see
eq. (\ref{mainformulamaintext}) below.

We validate the new formula (\ref{mainformulamaintext}) by reproducing in an efficient way the
known LO and NLO results. We then compute, for the first time, the EEC at NNLO.
The result is given in terms of an analytic  two-fold integral representation.
We present plots of the numerically integrated answer.

We also use the integral formula (\ref{mainformulamaintext}) in order to obtain limits of the energy correlator, 
namely the small angle  
and the back-to-back
 limits.  

  \medskip
 \noindent \textbf{2. EEC from correlation functions.}
 The detector operator that measures the energy flux in the direction $\vec n$ is given by
an integrated stress-energy tensor $T_{\mu\nu}$ \cite{Sveshnikov:1995vi,Korchemsky:1997sy,Korchemsky:1999kt,Belitsky:2001ij,Hofman:2008ar},
\begin{align}\label{definition-curlyE}
{\cal E}(\vec n)  =& \int_{- \infty}^{\infty} d \tau \ \lim_{r \to \infty} r^2  n^i T_{0i}(t=\tau+r, r \vec n) \, .
 \end{align}

The standard definition for the EEC as a differential cross-section can then be recast as
\begin{align}\label{def-EEC-main}
\EEC (\chi) =&  \int d \Omega_{\vec n_1} d \Omega_{\vec n_2} \delta(\vec n_1 \cdot \vec n_2 - \cos \chi ) \nonumber \\
& \hspace{-1cm} \times
\frac{\int d^4 x \e^{iqx} \vev{0|O^\dagger (x){\cal E}(\vec n_1){\cal E}(\vec n_2)O(0) |0}} { (q^0)^2 \int d^4 x \e^{iqx} \vev{0|O^\dagger (x) O(0) |0}} \,.
\end{align}
Here the operators $O$ (source) and $O^\dagger$ (sink) create the final state,  
whose particles are detected by the two calorimeters. 
Note that the operators are ordered as written, i.e. eq. (\ref{def-EEC-main}) expresses the EEC in terms of an integrated Wightman four-point correlation function.
The choice of the local operator $O$ depends on the physical problem. 
For  $e^+ e^-$ annihilation, $O$ is given by an electromagnetic current. 

{It is convenient to introduce the  variable
\begin{align} 
 \zeta=\frac{q^2 (n_1 \cdot n_2)}{2 (q\cdot n_1)(q \cdot n_2)} \, ,   
\end{align} 
where the null four-vector $n = (1, \vec n)$  characterizes the position of the detector. In   the rest frame of the source $q^\mu=(q^0,\vec 0)$,  we have $\zeta=\sin^2(\chi/2)$. In the same frame  the averaging over the angles  in (\ref{def-EEC-main}) becomes trivial  for a scalar source. 
}

In $\cN=4$ sYM, we may choose the source and sink 
to be scalar operators that are the bottom component of the supermultiplet of conserved currents.
As such, they are natural analogs of the electromagnetic current, and have fixed conformal weight two.
Moreover, the correlation function $\vev{0| OT_{\mu \nu}T_{\rho \sigma}O|0}$ can be obtained from the all-scalar correlation function $\vev{0| OOOO|0}$ by superconformal transformations. 
In this way, we can express the EEC in terms of  the four-point correlation function of {\it scalar operators} \cite{Belitsky:2013bja,Belitsky:2014zha,Korchemsky:2015ssa}.

Let us describe this relation in detail. We define the Euclidean correlation function
\begin{align}\label{def4ptcorrelator}
\vev{0|O (x_1) O(x_2) O(x_3) O (x_4)|0} =   \frac{\cG(z,\bz)}{(x_{13}^2 x_{24}^2)^2} \,,
\end{align}
depending on  the conformal cross-ratios
\begin{align}\label{definition_cross_ratios}
u=z\bz \equiv{x_{12}^2 x_{34}^2 \over x_{13}^2 x_{24}^2}
\, , \quad
v=(1-z)(1-\bz) \equiv{ x_{14}^2 x_{23}^2 \over x_{13}^2 x_{24}^2}   \,.
\end{align}
The  analytic continuation from the Euclidean to the Wightman function is easily obtained, for example, by using the Mellin transform of $\cG(u,v)$ \cite{Mack:2009mi,Belitsky:2013xxa,Belitsky:2013bja,Belitsky:2013ofa}. 
 Alternatively, we can convert the detector time integration in (1) to integration of certain discontinuities of $\mathcal{G}(u,v)$
\cite{Belitsky:2013xxa}. 
This approach is further improved by exploiting a different double discontinuity, first proposed in \cite{Caron-Huot:2017vep}, see also \cite{Simmons-Duffin:2017nub, Kravchuk:2018htv} . The latter is defined by 
\begin{align}
\text{dDisc}_{w=w_0}\,  g(w) = g(w) - \frac1{2} g (w^\lc ) - \frac1{2} g (w^\rc )\,,
\end{align}
where $w^\lc$ and $w^\rc$ refer to the mappings of the branch point $w=w_0$ to the two adjacent Riemann sheets. 
For example, $\text{dDisc}_{w=0}\, w^\a = 2\sin^2(\pi\a)\, w^\a$. 
In addition, in four dimensions and for a source of conformal weight two, the Fourier transform in \p{def-EEC-main} is equivalent to  another discontinuity \cite{Belitsky:2013xxa}. 
Putting all this together (for a detailed derivation see \cite{toappear}), 
we arrive at the {\it triple discontinuity} formula for the EEC 
\begin{align}\label{mainformulamaintext}
\EEC (\z ) & =  \frac{1}{ 4\pi^3 \z^2} \lim_{\ep\to0} \int_0^{1}  d \bz\,   \int_{0}^{\bz} d t \, \frac{1}{t(\z-\bz)+(1-\z)\bz}\nt 
& \times \text{dDisc}_{\bar z =1}  \text{Disc}_{z =0}    \left[(z-\bz)  \cG_\ep (z, \bar z)  \right]\,,    
\end{align}
where $0<\z<1$ \footnote{One expects also contact terms at the endpoint, 
which are required e.g. to verify the relation $\int_{-1}^{1}   \EEC (\chi) d\cos(\chi) = 1$ \cite{Basham:1978bw}.
In this letter, we assume the operators to be neither coincident nor back-to-back.}. After taking the discontinuities, the variable $z$ is treated as a function of the two  integration variables,
\begin{align} \label{defz}
z = \frac{\z t (t-\bz)}{t(\z-\bz)+(1-\z)\bz}\,.   
\end{align} 
The following comments are in order. 
In order to apply dDisc to power singularities, we employ an analytic regulator for the poles in $1-\bar{z}$,
\begin{align}\label{def-analytic-regulator}
\cG_\ep (z, \bar z)  \equiv  (1- \bar z )^{\ep}  \cG(z, \bar z) \,.
\end{align}
Instead of first computing the integrals in \p{mainformulamaintext} and taking the regulator $\ep\to0^+$ at the end, it is much more efficient to treat the integrand as a {\it singular distribution} \cite{Gelfand} of the `plus' type, $w^{-1+\ep}_+ =\frac1{\ep} \delta(w) +w_{+}^{-1} +\ep [w^{-1}\ln w]_++O(\ep^2)$. 
This allows us to drop many irrelevant terms, as illustrated in the one-loop example below.
Moreover, it allows us to automate the calculation at higher loops.

For convenience, we define
\begin{align}
 z \bar{z}  \Phi  \equiv (1-z) (1-\bar{z}) (z-\bar{z}) \cG \,,
 \end{align}
 and 
 \begin{align}\label{defF}
 F(\z) \equiv 4 \z^2 (1-\z) \EEC (\z) \,.
 \end{align}
 The correlation function $\cG$  has the perturbative expansion  
$\cG = \sum_{k \ge 0} a^k \cG^{(k)}$
 in the  `t Hooft coupling $a = N_c g^2_{\rm{YM}}/(4\pi^2) $. It 
was computed to two loops in \cite{Eden:2000mv,Bianchi:2000hn}, and at three loops in \cite{Eden:2011we,Drummond:2013nda}. 
Note that to obtain the EEC at N${}^{k}$LO, one needs $\cG$ at $(k+1)$ loops.
For any $\z \in  (0,1)$, we write the expansion
\begin{align}
F(\z; a) = a \,  F_{\rm LO} +a^2 F_{\rm NLO} + a^3 F_{\rm NNLO} + \mathcal{O}(a^4)\,.
\end{align}

    \medskip
 \noindent \textbf{3. EEC at LO.} 
At Born level $\Phi^{(0)} \sim z-\bz,$ and our triple discontinuity formula \p{mainformulamaintext} gives 0, which is the expected answer (up to contact terms). 
Let us now apply eq. \p{mainformulamaintext}  to the leading order (or one-loop) EEC, for which  \cite{Usyukina:1992jd} 
\begin{align} 
\Phi^{(1)} = \frac{1}{2}{\rm Li}_2(\bar z) - \frac{1}{2}{\rm Li}_2( z) + \frac{1}{4}\ln \lr{1- \bar z \over 1- z} \ln \lr{z\bar z}  \,.  
\end{align} 
The discontinuity at $z=0$ is
\begin{align}\label{lolo} 
\text{Disc}_{z =0}    \left[\frac{ \Phi^{(1)}}{(1- \bar z )^{1-\ep} } \right] =   \frac{ \pi}{4}  \frac{\ln (1-\bar z) - \ln(1- z)}{(1- \bar z )^{1-\ep}}\,.
\end{align}
The following operation $\text{dDisc}_{\bar z =1} $ is carried out keeping in mind the singular character of the distribution $w^{-1+\ep}$,
\begin{align}\label{dDisc_nologs}
\text{dDisc}_{w=0}\, w^{-1+\ep} &=2\sin^2(\pi\ep) \left\lbrack \frac{\delta(w)}{\ep}  +O(\ep^0)\right\rbrack  \,.
\end{align}
This expression vanishes in the limit $\ep\to0$. 
A non-trivial result is obtained due to the logarithms,  with the help of the identity 
\begin{align} \label{dDisc_Rules}
&\text{dDisc} [ w^{-1+\ep} \ln^m w  ] \ln^n w   \\
&=\partial^m_\ep \left[  2 \sin^2 (\pi \ep )\,\partial^n_\ep \left(\frac{\delta(w ) }{\ep } +  \sum_{k=0} \frac{\ep^k}{k!} \big[ w^{-1} \ln^k w  \big]_{+} \right)  \right]  \notag\,.
\end{align}
One such logarithm is already present in  eq. \p{lolo}. The $t-$integration in \p{mainformulamaintext} gives $A\ln(1-\bz)+ B$ with coefficients $A,B$ analytic at $\bz=1$. 
At LO we apply \p{dDisc_Rules} with $ m,n \leq 1$.  
 The operation dDisc  produces $ \delta(1-\bz)$, i.e. we observe a transcendental weight drop of three. The $\bz$ integral in \p{mainformulamaintext} is removed and we obtain, in accord with \cite{Engelund:2012re},
\begin{align}
F_{\rm LO}(\z) =-{\ln(1-\z)} \,.
\end{align}

 \medskip 
   \noindent \textbf{4. EEC at NLO and NNLO.}
The example above shows the way to higher perturbative orders. 

At NLO, $\Phi^{(2)}$ is given by so-called two-loop ladder and one-loop box squared integrals \cite{Eden:2000mv,Bianchi:2000hn}, expressed in terms of weight-four harmonic polylogarithms (HPL) \cite{Remiddi:1999ew}  in $z$ and $\bar z $.  
The HPL shuffle algebra allows us to extract all singular terms as $z\to0$ and $\bar{z} \to 1$.
This is conveniently implemented in the \textsc{Mathematica} package HPL \cite{Maitre:2005uu}.
In this way, we obtain powers of $ \ln z $ and $ \ln (1- \bar z)$, whose coefficients  are HPL 
that are analytic at $z= 0$ and $\bar z =1$. 
The {triple discontinuity} of the logarithms  lowers the transcendental weight  by three. 
We are left with a two-fold integration in $t$ and $\bar z $. 
We find that the integrand is linearly reducible \cite{Brown:2008um,Brown:2009ta,Bogner:2013tia,Panzer:2015ida}. (In the case of the one-loop box squared, we first change variables to achieve this.) As a consequence, we can carry out  all integrals algorithmically \cite{Bogner:2014mha,Panzer:2014caa}. This is conveniently done using the \textsc{Maple} program HyperInt  \cite{Panzer:2014caa}. The final result for the EEC  can be converted to classical polylogarithms and fully agrees  with \cite{Belitsky:2013ofa}. 
The symbol alphabet \cite{Goncharov:2010jf} is 
\begin{align}\label{symbolNLO}
\left\{ \zeta, 1-\zeta, \frac{1-\sqrt{\zeta}}{1+\sqrt{\zeta}} \right\}\,.
\end{align}

At NNLO, the integral formula  \p{mainformulamaintext}  exhibits some new algebraic features, 
due to the leading singularities (algebraic prefactors) of the three-loop integrals appearing in $\Phi^{(3)}$ and their symbol alphabet \cite{Drummond:2013nda}. 

The set of integrals are three-loop ladders,
 products of one- and two-loop boxes, as well 
as so-called easy and hard integrals that come in three different orientations. 
A new feature compared to two loops is that the hard integral is given in terms of 
a class of Goncharov polylogarithms (GPL) \cite{Goncharov:2010jf},
whose symbol contains an entry $z- \bar z $.  
We implement the GPL shuffle relations to extract the singularities at $z=0$ and $ \bar z=1$,
and take the triple discontinuity.
We find that all except two terms are linearly reducible and can be integrated,
giving HPL with transcendental weight ranging from two to five.
The symbol alphabet of these terms is the same as at NLO.
  
  \begin{figure}[t] 
\centerline{\includegraphics[height=55mm]{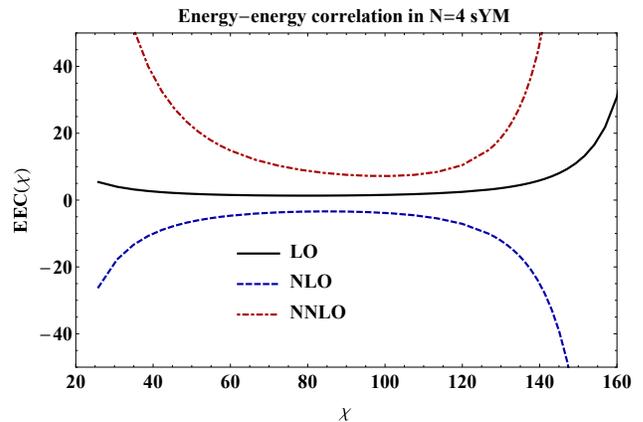}}
\caption{\small   EEC in $\mathcal{N}=4$ sYM. 
We display separately the contributions  from $F_{\rm LO}, F_{\rm NLO}$ and $ F_{\rm NNLO}$ to $\EEC(\chi)$ .  
} \label{fig2}
\end{figure}
 
For certain orientations, the easy and hard integrals have leading singularities that introduce a rational factor $1/(1-u)$ or $u/(u-v)$. 
These terms do not admit a change of integration variables that allows the  two-fold integration in \p{mainformulamaintext}  to be automated in HyperInt.   
We present our final three-loop result as a sum of polylogarithms $f_{\rm HPL}$, plus a two-fold finite integral,
 \begin{align}  \label{resultFatNNLO}
F_{\rm NNLO} (\zeta) &= f_{\rm HPL} (\zeta) +   \int_{0}^1 d \bar z  \int_{0}^{\bar z } dt \, \frac{\z-1}{t(\z-\bz)+(1-\z)\bz} \nt 
& \times \left[  R_1 ( z, \bar z )  P_1 (z, \bar z ) + 
  R_2 ( z, \bar z )  P_2 (z, \bar z ) \right] \,,
\end{align} 
 where 
 \begin{align} 
  R_1 = \frac{z \bar z }{ 1 - z - \bar z}\, , \quad  R_2 = \frac{z^2 \bar z }{ (1-z)^2  ( 1 - z \bar z)}   
 \end{align}
and $z$ is given by eq. \p{defz}. $P_{1}, P_2$ are HPL in $z, \bar z$ of weight three.
 We provide their explicit expressions, as well as the polylogarithm function $f_{\rm HPL} (\z)$ with symbol alphabet (\ref{symbolNLO}),  in the ancillary files. 
 
 After partial fractioning,   $R_1, R_2$ decompose into a sum of linearly reducible terms and  irreducible kernels $K_1, K_2$,  
 \begin{align} 
 K_1 & =  \frac{ \bar z (1- \bar z) \,  (\zeta -1)}{ (1-t) (1- \bar z ) \bar z + \z \,  (t-\bar z)(1- t  - \bar z )    }\,, \\ 
  K_2  & = \frac{1}{(1-t)^2 (1- \bar z )^2 } K_1 \left(  \frac{t}{t-1},  \frac{\bar z}{\bar z-1 } \right)\,.  \end{align}
Remarkably, upon the replacements $ \zeta \rightarrow \frac{m^2 }{ m^2 -S }$, $  t \rightarrow x_1+1 ,  \, \bar z  \rightarrow x_2 $,  the denominator of $K_1$ becomes the Symanzik polynomial of the equal mass sunrise integral \cite{Broadhurst:1987ei,Laporta:2004rb,Bloch:2013tra,Adams:2015ydq}. The features of the integrand that we observe here is evidence that the integral is elliptic.

 We evaluate numerically  the two-fold integrals in \p{resultFatNNLO} in \textsc{Mathematica} for 50 different values of $\z$ with precision $O (10^{-7})$.  In Fig.~\ref{fig2} we plot the EEC shape at the first three perturbative orders.

 \medskip 
\noindent \textbf{5.~ Back-to-back and collinear expansion.}
 The EEC is  logarithmically enhanced 
 in the extreme kinematic situations where the two detectors are back-to-back ($\z \rightarrow 1 $) or are close to each other ($ \z \rightarrow 0 $). 
We extract analytically the contribution of the two-fold integral in \p{resultFatNNLO} in both limits.

 In the back-to-back limit,  we observe 
 that as $\z \rightarrow 1$  the integrand as a function of $t$ and $\bar z$ is suppressed everywhere in the integration domain, hence the integral in \p{resultFatNNLO} only contributes subleading powers of $y=1-\zeta$.  The leading asymptotic behavior of $ F_\text{NNLO} (\z)$ is determined by  $f_{\rm HPL} (\z)$.
 At leading power in $y$,     
 \begin{align} \label{EECbacktoback}
 F_\text{NNLO} (\z)  \overset{ \z \rightarrow 1 }{\sim } &  -\frac18\, \ln^5 y  - \frac{\pi^2}{6}  \, \ln^3 y  -\frac{11}{4} \zeta_3 \,  \ln^2 y   \nt
&  - \frac{61}{ 720} \pi^4 \, \ln y   - \frac{\pi^2}{3}  \zeta_3 - \frac72 \zeta_5 \,.
 \end{align} 
 On the other hand, the $\z \rightarrow 1$ behavior of the EEC in $\mathcal{N}=4$ sYM is analogous to that in QCD. The Sudakov logarithms can be resummed to all loop orders \cite{Collins:1981uk}, 
 \begin{align}  \label{EECSudakov} 
 F(\z) \sim \frac{1}{2} H(a) \int _{0}^\infty db\, b J_0 (b) \, e^{-\frac12  \Gamma_\text{cusp} (a) L^2 - \Gamma(a) L}  \,.
 \end{align} 
Here $L  = \ln(  e^{ 2 \gamma_E} b^2 / 4y  )$; $J_{0}$ is a Bessel function; $\Gamma_\text{cusp} (a)$ and $\Gamma(a)$ are the  cusp and collinear anomalous dimensions, whose three-loop values can be found in \cite{Bern:2005iz}. 
 The function $H(a)$  describes hard emissions in the EEC. 

 We verify that our expression \p{EECbacktoback} agrees with the resummation  formula \p{EECSudakov} expanded to three-loop order. The coefficient of the single logarithm term allows us to determine the hard function up to NNLO, $ H(a) = 1- \zeta_2 \,a +5 \zeta_4 \, a^2$.\footnote{
The hard function $H(a)$ was independently obtained at two loops in \cite{privatecommGrisha}.} 

Remarkably, eq.~\p{EECbacktoback}  has homogeneous transcendental weight,  and so do the ingredients of the resummation formula. In fact, in the back-to-back limit,  we anticipate that $\mathcal{N} = 4$ sYM describes the maximally transcendental part of the QCD asymptotics   \cite{Belitsky:2013ofa}. This is confirmed by comparing with the recent QCD result \cite{Moult:2018jzp} upon the substitution $C_F, C_A \rightarrow N_c$.

 In the collinear limit, the elliptic integral in \p{resultFatNNLO} is not suppressed. 
 We extract its leading contribution by expanding in powers of $\z$ at the integrand level and integrating the leading term, which comes solely from the $R_1 \times P_1$ piece.  
 Combining with $f_{\rm HPL} (\zeta)$ in the $\z \rightarrow 0 $ limit, we obtain the three-loop small-angle asymptotics  
 \begin{align} 
 F_{\rm NNLO} (\z)   \overset{\z \rightarrow 0 }{\sim}  &\,  \z \left[ \frac{1}{2} \ln^2 \z +  \Big( - 5 +   \frac{\pi^2}{3}  -   \zeta_3  \Big)  \ln \z  \right. \nt 
 & \left. + 17 -  \frac{4\pi^2}{3}  - \zeta_3   + \frac{5\pi^4}{144} +\frac{3}{2}\,  \zeta_5 \right]\,.
 \end{align} 
A very nontrivial confirmation of our result is the agreement with the prediction of the light-ray OPE approach \cite{confcoll}. 
 
\begin{figure}[t]
\centerline{
\includegraphics[scale=0.3]{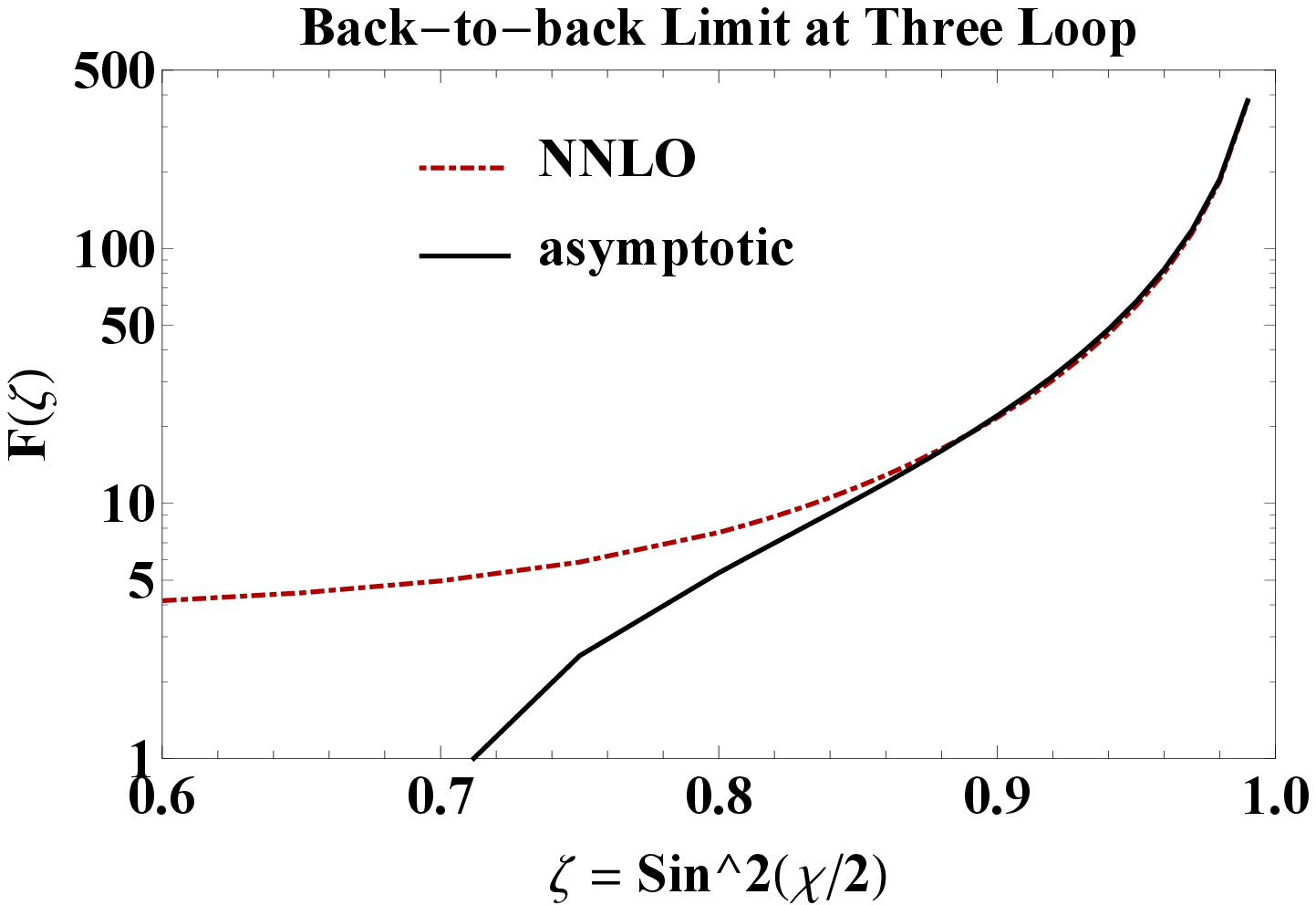}
\includegraphics[scale=0.3]{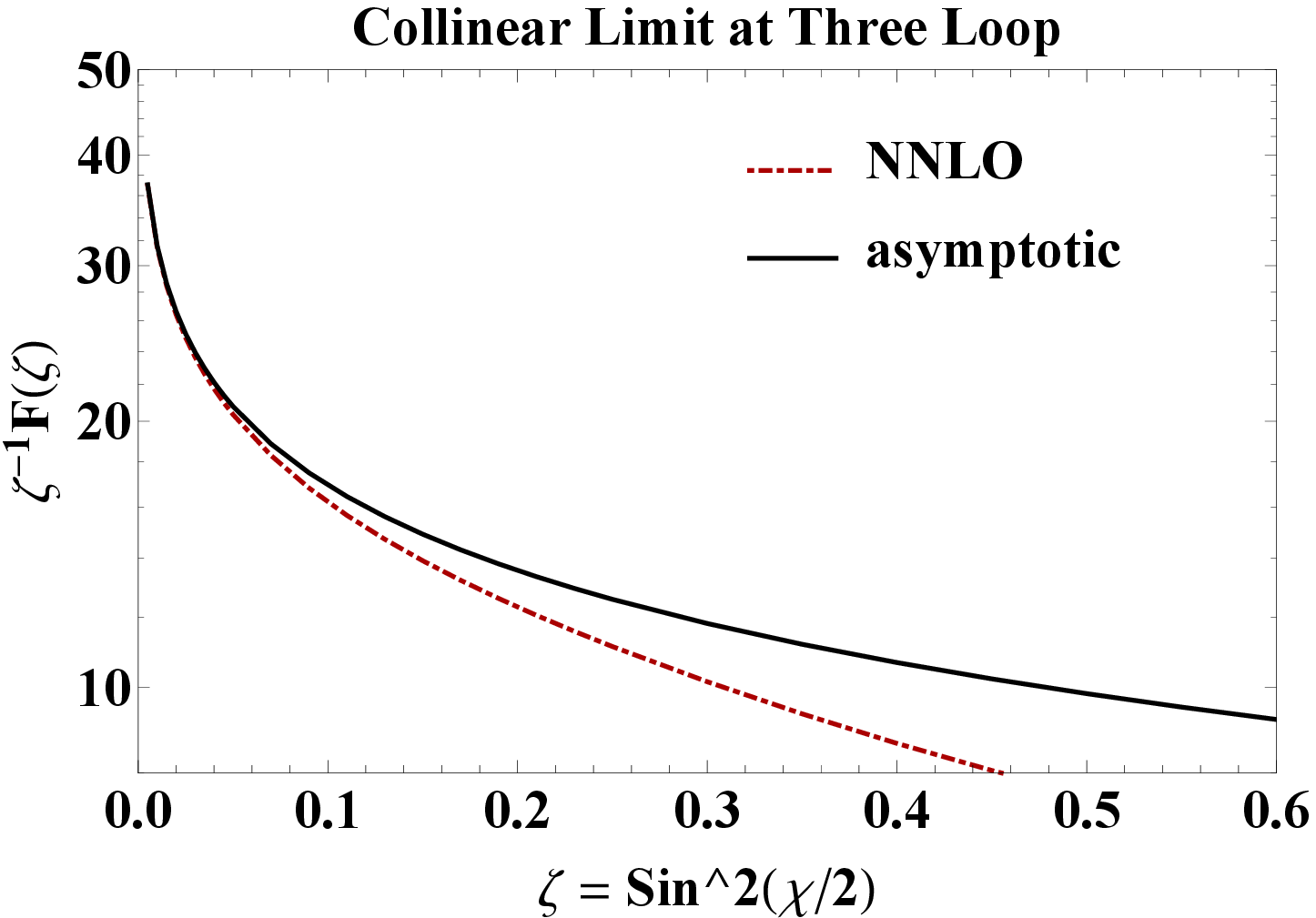}
}
\caption{\small Behavior of $F_{\rm NNLO}(\z)$ and its asymptotics in the back-to-back and collinear limits. 
}
\label{fig:EEClimits}
\end{figure}

In Fig.~\ref{fig:EEClimits}, we show the comparison between the leading power asymptotics and the numerical evaluation of (\ref{resultFatNNLO}).
 
  \medskip
\noindent \textbf{6. Outlook.} Our work opens the way for several applications. 
Our main formula (\ref{mainformulamaintext}) expresses the EEC in terms of a four-point function,
 a central object in any conformal field theory (CFT). The derivation of (\ref{mainformulamaintext}), 
to be given in \cite{toappear}, relied on the fact that the operator ${O}$ has fixed scaling dimension two. {It would be interesting to see  if a similar triple discontinuity formula exists  in a generic CFT, where the source and sink are replaced by a conserved current.} It would also be very interesting to explore how our formalism works in QCD.

We extracted the leading power asymptotics in the collinear and back-to-back limits at NNLO. 
We wish to emphasize that our result (\ref{resultFatNNLO}) contains information about subleading powers as well. The latter constitute useful data for resummations of large logarithms at subleading power \cite{Laenen:2008ux,Laenen:2010uz,Bonocore:2016awd,Boughezal:2016zws,Moult:2016fqy,Feige:2017zci,Moult:2017rpl,Moult:2017jsg,Moult:2018jjd,Balitsky:2017flc,Beneke:2017ztn,Beneke:2018gvs}.
Moreover, thanks to eq. (\ref{mainformulamaintext}) it may be possible to understand these limits, at arbitrary coupling, in terms of properties of the four-point correlation function.  

In $\mathcal{N}=4$ sYM, the integrand of the four-point function $\cG$ is known to ten loops \cite{Eden:2011we,Bourjaily:2016evz}. Once the integrated four-loop result becomes available \cite{Drummond:2006rz,Drummond:2012bg,Schnetz:2013hqa,Golz:2015rea,Caron-Huot:2014lda,Eden:2016dir}, our formula (\ref{mainformulamaintext}) can be used to obtain the EEC at N${}^{3}$LO.

We provided evidence that the EEC at NNLO contains elliptic functions. 
It suggests that the same type of function might also appear in QCD.
It would be interesting to fully work out the remaining integrals in (\ref{resultFatNNLO}), using multiple elliptic polylogarithms \cite{Adams:2016xah,Adams:2017ejb,Remiddi:2017har,Ablinger:2017bjx,Broedel:2018iwv,Broedel:2018qkq,Broedel:2019hyg}.

Finally, it would be interesting to apply our infrared finite approach to energy correlators with more than two detectors (cf. \cite{Moult:2016cvt,Komiske:2017aww,Chivukula:2017nvl} for related generalizations of energy correlations).
This may be important for the theoretical understanding of jet observables \cite{Coleman:2017fiq,Larkoski:2017jix,Asquith:2018igt,Komiske:2018cqr,Marzani:2019hun,Datta:2019ndh}.

\noindent \textbf{Acknowledgements.} We would like to thank D.~Chicherin for his generous help with the calculations and T.~Hahn,  M.~D.~Schwartz and G.~Zanderighi  for helpful conversations. We are grateful to the authors of Ref. \cite{Drummond:2013nda}, as well as to Ian Moult and Tong-Zhi Yang for useful correspondence. We acknowledge many enlightening discussions with G.~Korchemsky.  This work received funding from the European Research Council (ERC) under the European Union's Horizon 2020 research and innovation programme, {\it Novel structures in scattering amplitudes} (grant agreement No 725110). J.~M.~H., E.~S.~ and K.~Y. also wish to thank the Galileo Galilei Institute for hospitality during the workshop ``Amplitudes in the LHC era''. A.~Z. is grateful to M.~Kologlu, P.~Kravchuk, D.~Simmons-Duffin for many discussions on the subject. 

\bibliographystyle{apsrev4-1} 

\bibliography{joh_more_refs}

\end{document}